\documentclass[%
 aip,
rsi,%
 amsmath,amssymb,
reprint,%
floatfix,
]{revtex4-1}

\usepackage{graphicx}
\usepackage{braket}
\usepackage{bm}
\usepackage{algpseudocode}

\newcommand{\indi}[0]{I}
\newcommand{\indj}[0]{J}
\newcommand{\indmo}[0]{i}
\newcommand{\indmotwo}[0]{j}
\newcommand{\indmu}[0]{\mu}
\newcommand{\indnu}[0]{\nu}
\newcommand{\indx}[0]{a}
\newcommand{\indy}[0]{b}
\newcommand{\indz}[0]{c}
\newcommand{\then}[1]{N_\mathrm{#1}}
\newcommand{\thei}[0]{\mathrm{i}}
\newcommand{\theone}[0]{\mathbf{1}}
\newcommand{\indp}[0]{p}
\newcommand{\indq}[0]{q}
\newcommand{\indel}[0]{\alpha}
\newcommand{\indeltwo}[0]{\beta}
\newcommand{\theelcharge}[0]{e}

\newcommand{\theorig}[0]{\mathbf{O}}

\newcommand{\thet}[0]{\mathbf{T}}
\newcommand{\thetdot}[0]{\dot{\mathbf{T}}}

\newcommand{\ther}[1]{\mathbf{R}_{#1}}
\newcommand{\therdot}[1]{\dot{\mathbf{R}}_{#1}}
\newcommand{\therddot}[1]{\ddot{\mathbf{R}}_{#1}}
\newcommand{\thert}[0]{\mathbf{R}_\mathrm{T}}

\newcommand{\thenab}[1]{\boldsymbol{\nabla}_{#1}}

\newcommand{\theel}[1]{\mathbf{r}_{#1}}
\newcommand{\theelt}[0]{\mathbf{r}_\mathrm{T}}

\newcommand{\them}[1]{\ifthenelse{\equal{#1}{}}{\mathbf{M}}{M_{#1}}}
\newcommand{\thez}[1]{Z_{#1}}
\newcommand{\thea}[2]{\mathbf{A} (#1,#2)}

\newcommand{\theb}[1]{\ifthenelse{\equal{#1}{}}{\mathbf{B}}{B_\mathrm{#1}}}
\newcommand{\thebt}[1]{\ifthenelse{\equal{#1}{}}{\mathbf{\tilde{B}}}{\tilde{B}_{#1}}}

\newcommand{\thef}[2]{\mathbf{F}_{#1}^\mathrm{#2}}
\newcommand{\theop}[3]{\hat{#1}_\mathrm{#2}^\mathrm{#3}}
\newcommand{\thee}[2]{#1_\mathrm{#2}}

\newcommand{\theu}[2]{\hat{U}_{#1}^\mathrm{#2}}

\newcommand{\thewf}[3]{
    \ifthenelse{\equal{#1}{nuc}}
        {\ifthenelse{\equal{#3}{1}}{\psi_{#2} (\ther{},t)}{\psi_{#2}}}
        {\ifthenelse{\equal{#3}{1}}{\phi_{#2} (\mathbf{r};\ther{},\theorig)}{\phi_{#2}}}
}

\newcommand{\themo}[2]{
    \ifthenelse{\equal{#2}{1}}
        {\varphi_{#1} (\theel{};\ther{},\theorig{})}
        {\varphi_{#1}}
}

\newcommand{\theldn}[3]{\ifthenelse{\equal{#2}{1}}
    {\ifthenelse{\equal{#3}{1}} {\omega_{#1}^{*} (\theel{};\ther{},\theorig{})} {\omega_{#1} (\theel{};\ther{},\theorig{})}}
    {\ifthenelse{\equal{#3}{1}} {\omega_{#1}^{*}} {\omega_{#1}}}
}

\newcommand{\thegsn}[3]{\ifthenelse{\equal{#2}{1}}
    {\ifthenelse{\equal{#3}{1}} 
        {G_{#1}^{*} (\theel{};\ther{})} 
        {G_{#1} (\theel{};\ther{})}}
    {\ifthenelse{\equal{#3}{1}} 
        {G_{#1}^{*}} 
        {G_{#1}}}
}

\newcommand{\thelfac}[3]{\ifthenelse{\equal{#2}{1}}
    {\ifthenelse{\equal{#3}{1}}
        {g^* (\theel{},\ther{#1})}
        {g (\theel{},\ther{#1},\theorig{})}}
    {\ifthenelse{\equal{#3}{1}}
        {g_{#1}^*}
        {g_{#1}}}
}

\newcommand{\thecoef}[3]{\ifthenelse{\equal{#3}{1}}{C_{#1#2}^{*}}{C_{#1#2}}}

\newcommand{\thechi}[2]{\boldsymbol{\chi}_{#1} (#2)}
\newcommand{\thechit}[2]{\boldsymbol{\chi}_{#1}^\mathrm{T} (#2)}
\newcommand{\thedchi}[3]{\dfrac{\partial \boldsymbol{\chi}_{#1} (#3)}{\partial  \mathbf{R}_{#2}}}

\newcommand{\thefluc}[4]{\ifthenelse{\equal{#4}{0}}
{\boldsymbol{\alpha}_{#1#2} (#3)}
{[\boldsymbol{\alpha}_{#1#2} (#3)]^\mathrm{T}}
}

\newcommand{\themulop}[1]{\mathcal{M}_{#1}}

\newcommand{\theom}[3]{\boldsymbol{\Omega}_{#1#2} (#3)}
\newcommand{\theoms}[4]{\boldsymbol{\Omega}_{#1#2}^\mathrm{#4} (#3)}
\newcommand{\theomel}[3]{\Omega_{#1#2} (#3)}
\newcommand{\theomtotel}[3]{\Omega_\mathrm{tot}^{#1#2} (#3)}

\newcommand{\thedip}[3]{\boldsymbol{\mu}_{#1#2} (#3)}

\newcommand{\thegamma}[1]{\boldsymbol{\gamma}_{#1}}
\newcommand{\thegammael}[2]{\gamma_{#1}^{#2}}

\newcommand{\theqm}[3]{Q_{#1#2}^\mathrm{M} (#3)}

\newcommand{\thepelop}[1]{\mathbf{\hat{p}}_{#1}}
\newcommand{\thepielop}[1]{\boldsymbol{\hat{\pi}}_{#1}}
\newcommand{\thepieldir}[2]{\hat{\pi}_{#1}^{#2}}
\newcommand{\thekelop}[1]{\mathbf{\hat{k}}_{#1}}
\newcommand{\thekeldir}[2]{\hat{k}_{#1}^{#2}}

\newcommand{\thepnucop}[1]{\mathbf{\hat{P}}_{#1}}
\newcommand{\thepnucopu}[1]{\mathbf{\hat{P}}_{#1}^\mathrm{U}}
\newcommand{\thepinucop}[1]{\boldsymbol{\hat{\Pi}}_{#1}}
\newcommand{\thepnucdir}[2]{\hat{P}_{#1}^{#2}}

\begin{document}

\title[]{Magnetic-Translational Sum Rule and Approximate Models\\ of the Molecular Berry Curvature}

\author{Laurens D. M. Peters}
\email{laurens.peters@kjemi.uio.no}
\affiliation
{Hylleraas Centre for Quantum Molecular Sciences,  Department of Chemistry, 
University of Oslo, P.O. Box 1033 Blindern, N-0315 Oslo, Norway}
\author{Tanner Culpitt}
\affiliation
{Hylleraas Centre for Quantum Molecular Sciences,  Department of Chemistry, 
University of Oslo, P.O. Box 1033 Blindern, N-0315 Oslo, Norway}
\author{Erik I. Tellgren}
\affiliation
{Hylleraas Centre for Quantum Molecular Sciences,  Department of Chemistry, 
University of Oslo, P.O. Box 1033 Blindern, N-0315 Oslo, Norway}
\author{Trygve Helgaker}
\affiliation
{Hylleraas Centre for Quantum Molecular Sciences,  Department of Chemistry, 
University of Oslo, P.O. Box 1033 Blindern, N-0315 Oslo, Norway}

\date{\today}

\begin{abstract}
The Berry connection and curvature are key components of electronic structure calculations for atoms and molecules in magnetic fields. They ensure the correct translational behavior of the effective nuclear Hamiltonian and the correct center-of-mass motion during molecular dynamics in these environments. In this work, we demonstrate how these properties of the Berry connection and curvature arise from the translational symmetry of the electronic wave function and how they are fully captured by a finite basis set of London orbitals but not by standard Gaussian basis sets. This is illustrated by a series of Hartree--Fock calculations  on small molecules in different basis sets. Based on the resulting physical interpretation of the Berry curvature as the shielding of the nuclei by the electrons, we introduce and test a series of approximations using the Mulliken fragmentation scheme of the electron density. These approximations will be particularly useful in \textit{ab initio} molecular dynamics calculations in  a magnetic field, since they reduce the computational cost, while recovering the correct physics and up to 95\% of the exact Berry curvature.
\end{abstract}

\maketitle

\section{Introduction}

The Berry connection and the Berry curvature can be interpreted, respectively, as the vector potential and field related to a geometric (or Berry) phase.\cite{Berry1984,Mead1992,Anandan1997,Resta2000} Therefore, these quantities are important concepts in many fields of modern physics (for example, the physics of crystals\cite{Zak1989} or conical intersections\cite{Mead1979}) and play a key role in understanding the quantum Hall\cite{Prange1990} and the Aharonov--Bohm effect\cite{Aharonov1959}. 

Less widely known are the implications of the Berry phase for molecules in a magnetic field. Schmelcher and Cederbaum\cite{Schmelcher1988,Schmelcher1989,Schmelcher1997} and later Yin and Mead\cite{Yin1992,Yin1994} as well as Peternelj and Kranjc\cite{Peternelj1993} reported that the Born--Oppenheimer approximation in a magnetic field gives rise to a geometric vector potential (or Berry connection) in the effective nuclear Hamiltonian, resulting in an additional velocity-dependent force in the nuclear equations of motion that depends on a Berry curvature [$\boldsymbol{\Omega}_{\indi\indj}$],\cite{Ceresoli2007}
\begin{align}
\thef{\indi}{B}{} &=
\sum \limits_{\indj=1}^{\then{nuc}} \boldsymbol{\Omega}_{\indi\indj}  \therdot{\indj} ,
\label{int_000}
\end{align}
where $\therdot{\indj}$ is the velocity of nucleus $\indj$. This Berry force is essential for the correct physics of molecules in a magnetic field as it describes the screening of the nuclei by the electrons.\cite{Yin1992,Schmelcher1997,Ceresoli2007} Neglecting the Berry force in molecular dynamics is thus equivalent to omitting the electrons when calculating the Lorentz force acting on the nuclei.  

Since these conceptual discussions almost two decades ago, the field of electronic-structure calculations in a magnetic field has grown rapidly, allowing for predictions at magnetic field strengths up to $B_0 = 2.35 \times 10^{5}\,\mathrm{T}$. Besides numerical methods\cite{Ozaki1993,Kravchenko1996,Ivanov1998,Jones1999,Ivanov1999,Ivanov2000,Thirumalai2009,Thirumalai2014,Lehtola2020} and anisotropic Gaussian functions\cite{Schmelcher1988a,Kappes1994,Detmer1997,Detmer1998,Al-Hujaj2000,Becken2002,Al-Hujaj2004}, London atomic orbitals\cite{London1937,Hameka1958,Ditchfield1976,Helgaker1991} have become popular, combining standard Gaussian basis sets with London phase factors that include a field dependence and gauge-origin dependence. The advantage of the London-orbital approach is that it can be combined with various quantum-chemistry methods (for example, Hartree--Fock (HF) theory, density-functional theory, and coupled-cluster theory) to calculate energies and properties of molecular systems in a magnetic field.\cite{Tellgren2008,Tellgren2009,Lange2012,Tellgren2012,Reynolds2015,Stopkowicz2015,Hampe2017,Irons2017,Hampe2019,Sen2019,Sun2019,Austad2020,Hampe2020,Pausch2020,Williams-Young2020,Irons2021,Blaschke2022}

Very recently, Culpitt \emph{et al.}\ derived and presented a numerical scheme for calculating the Berry curvature in a magnetic field at the HF level of theory using London orbitals.\cite{Culpitt2021} Their approach was later extended to an analytical calculation\cite{Culpitt2022} and used to conduct the first extensive study of \textit{ab initio} molecular dynamics in a magnetic field.\cite{Peters2021,Monzel2022} While these publications confirmed the properties of the Berry curvature discussed previously,\cite{Yin1992,Schmelcher1997,Ceresoli2007} it remained unclear why certain properties of the exact wave function are reproduced exactly only in the basis-set limit or in a finite basis set of London orbitals. 

In this work, we re-examine the geometric vector potential and the Berry curvature in a magnetic field. In particular, we are interested in the fulfillment of the \textit{magnetic-translational sum rule}
\begin{equation}
\sum \limits_{\indi,\indj=1}^{\then{nuc}} \boldsymbol{\Omega}_{\indi\indj} 
= \theelcharge{} \then{el} \thebt{},
\label{int_001}
\end{equation}
where $e$ is the unit charge, $\then{el}$ is the number of electrons, and $\thebt{}$ is the magnetic field tensor, related to the magnetic field $\theb{}$ by $\tilde B_{ab} = - \epsilon_{abc}B_c$ where $\epsilon_{abc}$ is the Levi-Civita symbol. In a magnetic field, the correct center-of-mass motion is obtained only if this relation is satisfied exactly.

In this paper, we demonstrate how eq.\,\eqref{int_001} follows from the translational symmetry of the exact electronic wave function and also how it is captured by approximate wave functions in a complete one-electron basis or in a finite basis of London orbitals. Additionally, we present approximations to $\boldsymbol{\Omega}_{\indi\indj}$ that fulfill eq.~\eqref{int_001} while reducing computational cost. For more details on the implementation and calculation of the geometric vector potential and the Berry curvature, the reader is referred to refs.~\onlinecite{Culpitt2021} and \onlinecite{Culpitt2022}.

The theory section (Section~II) is organized as follows. After reviewing notation in Section~\ref{notation} and the electronic Hamiltonian in a magnetic field in Section~\ref{elham}, we discuss the nuclear equations of motion in a magnetic field in Section\;\ref{NEM}, with emphasis on the Berry force and Berry curvature. Having introduced the electronic pseudomomentum in Section\;\ref{ElPseudo}, we investigate the behavior of the Hamiltonian and the wave function upon translation of the nuclei in Section\;\ref{NucHamWF} and the corresponding behavior of the total nuclear canonical momentum in Section\;\ref{secUPU}. With this information at hand, we discuss the total Berry curvature and total Berry connection in Sections~\ref{BCurvature} and~\ref{BConnection}, respectively, both for exact wave functions. The total Berry curvature in HF theory is then discussed in Sections\;\ref{HartreeFock} and \ref{HartreeFockLDN}, treating both exact HF theory and HF theory using a finite basis of London atomic orbitals. Finally, Section~\ref{Mulliken} presents a series of approximate Berry curvatures based on the Mulliken approach\cite{Mulliken1955} for atomic charges. We summarize computational details in Section\;III and illustrate our theoretical findings by calculations on various organic molecules at the HF level of theory in Section~IV, followed conclusions in Section~V.

\section{Theory}

\subsection{Notation}
\label{notation}

We use indices $\indx, \indy, ...$ for the Cartesian components $x$, $y$, and $z$, indices $\indel,\indeltwo, ...$ for the $\then{el}$ electrons, indices $\indi, \indj, ...$ for the $\then{nuc}$ nuclei, indices $\indmo, \indmotwo, ...$ for the $\then{occ}$ occupied molecular orbitals (MOs), indices $\indp, \indq, ..$ for the $\then{orb}$ general MOs, and indices $\indmu, \indnu, ...$ for the $\then{bas}$ atomic basis functions.
 
 We use $\theelcharge{}\thez{\indi}$, $\them{\indi}$, $\ther{\indi}$, and $\therdot{\indi}$ to represent the charge, mass, coordinates, and velocity of nucleus $\indi$, respectively, while $\theel{\indel}$ and $-\theelcharge{}$ are the coordinates and charge of electron $\indel$, respectively. 
 The electronic and nuclear coordinates are collectively denoted by
$\theel{}$ and $\ther{}$, respectively, while we use
\begin{align}
\ther{} + \thert{} = \begin{pmatrix}\ther{1}+\thet{} \\ \ther{2}+\thet{} \\ \vdots \end{pmatrix}, \quad
\theel{} + \theelt{} = \begin{pmatrix}\theel{1}+\thet{} \\ \theel{2}+\thet{} \\ \vdots \end{pmatrix}
\label{not_002}
\end{align}
to denote the translation of all nuclei and electrons of the system, respectively, by the amount $\thet{}$.

\subsection{Electronic Hamiltonian and Wave Function}
\label{elham}

Consider a molecular electronic system of $\then{el}$ electrons in a uniform magnetic field $\theb{}$, which is represented by a vector potential in the Coulomb gauge
\begin{equation}
\thea{\mathbf{u}}{\theorig{}} = 
\dfrac{1}{2} \theb{} \times (\mathbf{u} - \theorig{}) \label{Avec}
\end{equation}
with gauge origin $\theorig{}$. At a given nuclear geometry, $\ther{}$ the electronic Hamiltonian without the spin-Zeeman term is given by
\begin{align}
\theop{H}{el}{} (\theel{},\ther{},\theorig{}) 
&=  \dfrac{1}{2} \sum_{\indel=1}^{\then{el}} 
\thepielop{\indel}^2 (\theel{},\theorig{}) + \theop{V}{el}{} (\theel{};\ther{}) ,
\label{ham_002}
\end{align}
where $\theop{V}{el}{} (\theel{};\ther{})$ is the external potential operator and $\thepielop{\indel} (\theel{}, \theorig{})$ is the kinetic momentum of electron $\indel$, which in terms of the canonical momentum
\begin{align}
\thepelop{\indel{}}  &= - \thei \hbar \thenab{\indel}, \label{not_003} 
\end{align}
and the vector potential in eq.\,\eqref{Avec} takes the form
\begin{align}
\thepielop{\indel{}} (\theel{},\theorig{}) 
&= \thepelop{\indel{}} + 
\theelcharge{} \thea{\theel{\indel{}}}{\theorig{}} .
\label{not_004} 
\end{align}
We assume that the external potential $\theop{V}{el}{} (\theel{};\ther{})$ is translationally invariant, depending only on relative coordinates.

We denote the electronic ground-state wave function by $\thewf{el}{}{0}(\theel{}; \ther{}, \theorig{})$, which in the exact case is an eigenfunction of the electronic Hamiltonian:
\begin{equation}
\theop{H}{el}{} (\theel{},\ther{},\theorig{})\thewf{el}{}{0}(\theel{}; \ther{}, \theorig{}) =  \thee{E}{BO}(\ther{}) \thewf{el}{}{0}(\theel{}; \ther{}, \theorig{}),
\label{schrodinger}
\end{equation}
where the eigenvalue $\thee{E}{BO}(\ther{})$ is the Born--Oppenheimer ground-state energy. If the wave function is not an eigenstate of the Hamiltonian, then the Born--Oppenheimer energy can be calculated as an expectation value,
\begin{align}
\thee{E}{BO}(\ther{}) \!=\! 
\Braket{\thewf{el}{}{0} (\ther{},\theorig{})|\theop{H}{el}{} (\ther{},\theorig{})|\thewf{el}{}{0} (\ther{},\theorig{})} .
\label{ham_003}
\end{align}
The ground-state energy depends on the nuclear geometry but -- for
the exact wave function and for certain approximate wave functions to be considered here -- not on the gauge origin.

\subsection{Nuclear Equations of Motion}
\label{NEM}

The effective nuclear Born--Oppenheimer Hamiltonian in a uniform magnetic field is given by
\cite{Schmelcher1988,Schmelcher1989,Schmelcher1997,Yin1992,Yin1994,Ceresoli2007,Culpitt2021}
\begin{align}
\theop{H}{eff}{} (\ther{},\theorig{}) = 
\sum_{\indi=1}^{\then{nuc}} \dfrac{1}{2 \them{\indi}} 
&\bigl[
\thepinucop{\indi} (\ther{},\theorig{}) + \thechi{\indi}{\ther{},\theorig{}} 
\bigr]^2 \nonumber \\
&\qquad\qquad\qquad + \thee{E}{BO}(\ther{}) .
\label{ham_000}
\end{align}
It consists of the kinetic momentum operator and the geometric vector potential (Berry connection), respectively, of nucleus $\indi$
\begin{align}
\thepinucop{\indi{}} (\ther{},\theorig{}) 
&= \thepnucop{\indi{}} - \theelcharge{} \thez{\indi} \thea{\ther{\indi}}{\theorig{}} \label{not_007} 
\\
\thechi{\indi}{\ther{},\theorig{}} &= 
\Braket{\thewf{el}{}{0}(\ther{},\theorig{})| \thepnucop{\indi} \thewf{el}{}{0}(\ther{},\theorig{})} ,
\label{chiI}
\end{align}
where the canonical momentum operator of nucleus $\indi$ is given by
\begin{align}
\thepnucop{\indi{}} &= 
- \thei \hbar \thenab{\indi}. \label{not_006}
\end{align}
In setting up the Hamiltonian in eq.~\eqref{ham_000}, we have omitted the usually small diagonal Born--Oppenheimer correction (DBOC) to the Born--Oppenheimer potential.

\subsubsection{Classical Equations of Motion and Forces}

Taking $\theop{H}{eff}{} (\ther{},\theorig{})$ as the starting point, we may set up the classical nuclear equations of motion\cite{Schmelcher1988,Schmelcher1989,Schmelcher1997,Ceresoli2007,Culpitt2021,Peters2021}
\begin{align}
\them{\indi} \therddot{\indi}  &=   \thef{\indi}{}{(\ther{},\therdot{})}
\label{ham_004}
\end{align}
with three contributions to the force on atom $\indi$:
\begin{align}
\thef{\indi}{}{}
&= \thef{\indi}{BO}{} (\ther{}) + \thef{\indi}{L}{} (\therdot{})  + \thef{\indi}{B}{} (\ther{},\therdot{})  \nonumber \\
&=
 - \thenab{\indi} \thee{E}{BO}(\ther{}) -
\thez{\indi} \theelcharge{}  \theb{} \times \therdot{\indi} 
+ \sum \limits_{\indj=1}^{\then{nuc}} \theom{\indi}{\indj}{\ther{}}  \therdot{\indj}
\label{ham_005}
\end{align}
While the first contribution to the force is the familiar position-dependent Born--Oppenheimer force and the second contribution is the velocity-dependent Lorentz force, the third contribution depends on both the position and the velocity of the nucleus and is known as the Berry force. Without the Berry force, each nucleus would experience the applied external field rather than the local field resulting from the screening of the electrons.

The Berry force is obtained from the Berry curvature $\theom{\indi}{\indj}{\ther{}}$, which in turn is related to the Berry connection as
\begin{align}
\theom{\indi}{\indj}{\ther{}} &=
\thedchi{\indi}{\indj}{\ther{},\theorig{}} - \bigg[ \thedchi{\indj}{\indi}{\ther{},\theorig{}} \bigg]^\mathrm{T},
\label{ham_006a}
\end{align}
where we have introduced the Jacobian
\begin{equation}
\thedchi{\indi}{\indj}{\ther{},\theorig{}} = \left[ \thenab{\indj} \thechit{\indi}{\ther{},\theorig{}} \right]^\mathrm T,
\label{chiRR}
\end{equation}
and used the fact that the origin dependence of the Berry curvature vanishes (as shown in ref.~\onlinecite{Culpitt2021}) for the exact wave function and for certain approximate wave functions to be considered here.

From eq.~\eqref{chiI} and eq.~\eqref{ham_006a} we obtain the following expression for the Berry curvature in terms of the electronic wave function:
\begin{align}
\theom{\indi}{\indj}{\ther{}} 
&= \dfrac{\thei}{\hbar}\Braket{\thepnucop{\indi} \thewf{el}{}{0}(\ther{},\theorig{})| \thepnucop{\indj}^\mathrm{T} \thewf{el}{}{0}(\ther{},\theorig{})}
\nonumber \\
&\quad - \dfrac{\thei}{\hbar}
\Braket{\thepnucop{\indj} \thewf{el}{}{0}(\ther{},\theorig{})| \thepnucop{\indi}^\mathrm{T} \thewf{el}{}{0}(\ther{},\theorig{})}^\mathrm{T}
\label{ham_006b}
\end{align}
The Berry curvature is thus a second-order nonadiabatic matrix element, which provides the screening of the magnetic field acting on the nuclei by the electrons in the system.\cite{Schmelcher1988,Schmelcher1989,Schmelcher1997,Yin1992,Yin1994,Peternelj1993,Ceresoli2007,Culpitt2021,Peters2021} 

\subsubsection{Total Berry Curvature}

Before turning to the general case, let us  consider a neutral molecule 
in a uniform magnetic field. For a rigid system moving with a constant velocity $\thetdot{}$, the total force acting on all atoms adds up to zero:
\begin{equation}
\sum_{\indi=1}^{\then{nuc}} \thef{\indi}{}{} = \mathbf 0.
\end{equation}
Since $\sum_{\indi=1}^{\then{nuc}} \thef{\indi}{BO}{} = \mathbf 0$ follows from the translational symmetry of the Born--Oppenheimer potential energy $\thee{E}{BO}(\ther{})$, the total Lorentz and Berry forces must cancel in this case. From eq.~\eqref{ham_005}, we then obtain
\begin{equation}
\theelcharge{} \then{el} \theb{} \times \thetdot{} = \theom{\mathrm{tot}}{}{\ther{}} \thetdot{} \label{BTOmegaT}
\end{equation}
as a requirement for this cancellation. We have here used the fact that $\sum_{\indi=1}^{\then{nuc}} \thez{\indi} = \then{el}$ for a neutral system and introduced the total Berry curvature as the sum over all its $3 \times 3$ blocks:
\begin{align}
\theom{\mathrm{tot}}{}{\ther{}} &=
\sum_{\indi,\indj=1}^{\then{nuc}} 
\theom{\indi}{\indj}{\ther{}} .
\label{ham_008}
\end{align}
To eliminate the velocity from eq.~\eqref{BTOmegaT}, we represent $\theb{}$ in terms of the three-by-three magnetic field tensor, whose elements are related to the Cartesian field components 
$\thebt{\indx\indy} = - \epsilon_{\indx\indy\indz}\theb{\indz}$:
\begin{align}
\thebt{} =
\begin{pmatrix} 
0 & - \theb{z} & \theb{y} \\
\theb{z} & 0 & -\theb{x} \\
-\theb{y} & \theb{x} & 0
\end{pmatrix} .
\label{not_000}
\end{align}
Since $\thebt{} \thetdot{} = \theb{} \times \thetdot{}$, we may write the relation in eq.\,\eqref{BTOmegaT}
\begin{equation}
\theom{\mathrm{tot}}{}{\ther{}} = \theelcharge{} \then{el} \thebt{} ,
\label{mtsr}
\end{equation}
showing that the total Berry curvature is independent of the nuclear coordinates of the molecule and that it determines the magnetic field completely. From now on, we will refer to eq.~\eqref{mtsr} as the magnetic-translational sum rule. It is essential for the correct center-of-mass motion of a general molecule, stating that an overall translation does not induce a force on the center of mass of the neutral molecule, while the force on the center of mass of a charged molecule matches the Lorentz force of its total charge. We emphasize, however, that rotations and vibrations may affect the center-of-mass motion of a nonrigid molecule, even if the molecule is neutral.

In the following, we demonstrate that the magnetic-translational sum rule holds for the exact electronic wave function, for the exact HF wave function, and for HF wave functions expanded in a finite basis of London atomic orbitals. 

\subsection{Total Electronic Pseudomomentum}
\label{ElPseudo}

Of interest to us here is the pseudomomentum of the electrons within the Born--Oppenheimer approximation. The pseudomomentum operator of electron $\indel$ in a magnetic field $\theb{}$ is defined as\cite{Johnson1983,Schmelcher1988,Schmelcher1989}
\begin{align}
\thekelop{\indel{}} (\theel{},\theorig{}) 
&= \thepielop{\indel{}} - \theelcharge{} \theb{} \times \theel{\indel{}},
\label{not_005x}
\end{align}
which in the Coulomb gauge of eq.~\eqref{Avec} takes the form
\begin{align}
\thekelop{\indel{}} (\theel{},\theorig{}) 
= \thepelop{\indel{}} - \dfrac{\theelcharge{}}{2} \theb{} \times (\theel{\indel{}} + \theorig).
\label{not_005}
\end{align}
We note the following commutators between the Cartesian components (denoted by $\indx$, $\indy$, and $\indz$) of the pseudomomentum and kinetic momentum operators of electron $\indel$:\cite{Schmelcher1989}
\begin{align}
[\thekeldir{\indel}{\indx}, \thekeldir{\indel}{\indy}] &= 
-[\thepieldir{\indel}{\indx}, \thepieldir{\indel}{\indy}] = 
\thei \hbar \theelcharge \epsilon_{\indx\indy\indz} \theb{\indz} \label{comkk},\\
[\thekeldir{\indel}{\indx}, \thepieldir{\indel}{\indy} ] &= 0,\label{comkpi}
\end{align}
where we have omitted arguments for clarity. Introducing the operators for the total pseudomomentum of the electrons and the total canonical momentum of the nuclei, respectively,
\begin{align}
\thekelop{\mathrm{tot}} = \sum \limits_{\indel{} = 1}^{\then{el}} \thekelop{\indel} , \quad
\thepnucop{\mathrm{tot}} &= \sum_{\indi=1}^{\then{nuc}} \thepnucop{\indi} ,
\end{align}
and using the commutation relation in eq.~\eqref{comkpi}, we obtain
 \begin{align}
\left[ \theop{H}{el}{}, \thekelop{\mathrm{tot}} \right] &=
\thei \hbar\sum_{\indel=1}^{\then{el}}\dfrac{\partial \theop{V}{el}{} (\ther{})}{\partial \theel{\indel}}
 = -  \thei \hbar \sum_{\indi=1}^{\then{nuc}} \dfrac{\partial \theop{V}{el}{} (\ther{})}{\partial \ther{\indi}} 
 \nonumber \\
 &= - \left[ 
 \theop{V}{el}{} (\ther{}), \thepnucop{\mathrm{tot}}
 \right].
 \end{align}
It follows that $\thekelop{\mathrm{tot}} + \thepnucop{\mathrm{tot}}$ commutes with the electronic Hamiltonian,
\begin{align}
\left[ \theop{H}{el}{}, \thekelop{\mathrm{tot}} + \thepnucop{\mathrm{tot}}\right] &= 0,
\end{align}
and hence that the sum of the total nuclear canonical momentum and the total electronic pseudomomentum constitutes a constant of motion for the electrons in the presence of a magnetic field, within the Born--Oppenheimer approximation.\cite{Johnson1983,Schmelcher1988,Schmelcher1989}

\subsection{Nuclear-translated Hamiltonian and Wave Function}
\label{NucHamWF}

We now introduce the unitary operator
\begin{align}
\theu{}{}(\theel{}, \ther{},\theorig{},\thet{}) &= 
\exp\left\{ 
\theop{K}{}{} (\theel{}, \ther{},\theorig{},\thet{})
\right\} 
\nonumber \\
&=
\exp\left\{ - \dfrac{\thei}{\hbar} \thet{} \cdot \thekelop{\mathrm{tot}} + \dfrac{\thei}{\hbar} \eta(\ther{}, \theorig{}, \thet{}) \right\} \label{Uop}
\end{align}
using the total pseudomomentum operator of the electrons defined above and $\eta(\ther{}, \theorig{}, \thet{})$ as a real-valued, differentiable gauge function. Since the latter can take many different forms, there exists a set of $\theu{}{}$'s differing by a phase factor. Using the fact the components in the exponential of eq.~\eqref{Uop} commute
\begin{equation}
\left[\thet{} \cdot \thepelop{\indel} ,  
\thet{} \cdot \dfrac{\theelcharge{}}{2} \theb{} \times (\theel{\indel} + \theorig{})
\right] 
= \dfrac{\thei \theelcharge}{2 \hbar} \theb{} \cdot (\thet{} \times \thet{}) = 0,
\end{equation}
we may factorize the unitary operator in the form
\begin{equation}
\theu{}{}(\theel{}, \ther{}, \theorig{},\thet{}) = 
\theop{f}{}{}(\ther{}, \theorig{}, \thet{}) \theop{g}{}{}(\theel{}, \thet{}) \theop{t}{}{}(\thet{}) \label{fgt}
\end{equation}
where we have introduced the unitary operators
\begin{align}
\theop{g}{}{}(\theel{}, \thet{}) &= \exp\left\{ - \dfrac{\thei \theelcharge}{2\hbar} ( \theb{} \times \thet{} ) \cdot \theel{\mathrm{tot}}  \right\} , \label{Gop}\\
\theop{t}{}{}(\thet{}) &= \exp\left\{ - \dfrac{\thei}{\hbar} \thet{} \cdot \thepelop{\mathrm{tot}} \right\} \label{Top} ,\\
\theop{f}{}{}(\ther{}, \theorig{}, \thet{}) &= \exp\left\{\dfrac{\thei}{\hbar} \eta(\ther{}, \theorig{}, \thet{})
- \dfrac{\thei \theelcharge}{2\hbar} \textstyle \sum_\indel{} ( \theb{} \times \thet{} ) \cdot \theorig{}
\right\} ,
\end{align}
in terms of
\begin{equation}
\theel{\mathrm{tot}}= \sum_{\indel=1}^{\then{el}}\theel{\indel}, \quad 
\thepelop{\mathrm{tot}}=
\sum_{\indel=1}^{\then{el}}\thepelop{\indel}.
\end{equation}
We note that $\theop{t}{}{}(\thet{})$ and $\theop{g}{}{}(\theel{}, \thet{})$ translate the coordinates of every electron and the gauge origin by a vector $-\thet{}$, respectively, while $\theop{f}{}{}(\ther{}, \theorig{}, \thet{})$ is an arbitrary phase factor. The order of the operators in eq.~\eqref{fgt} does not matter since their arguments commute with one another.

Consider now the unitary transformation of the Hamiltonian by the operator $\theu{}{} (\theel{}, \ther{},\theorig{},\thet{})$. By translation symmetry, we have
\begin{equation}
\theop{H}{el}{}(\theel{}, \ther{}, \theorig{})
= \theop{H}{el}{}(\theel{}+ \theelt{}, \ther{}+ \thert{}, \theorig{} + \thet{}) \label{Htra}
\end{equation}
Omitting the arguments of $\hat{t}$, $\hat{g}$, and $\hat{f}$ for brevity, we then obtain
\begin{align}
&\theu{}{} (\theel{}, \ther{},\theorig{},\thet{}) \theop{H}{el}{}(\theel{}, \ther{}, \theorig{}) \theu{}{\dagger} (\theel{}, \ther{},\theorig{},\thet{}) \nonumber \\
&=\theop{f}{}{} \theop{g}{}{} \theop{t}{}{} \theop{H}{el}{}(\theel{}+ \theelt{}, \ther{}+ \thert{}, \theorig{} + \thet{}) \theop{t}{}{\dagger} \theop{g}{}{\dagger} \theop{f}{}{\dagger}
\nonumber \\
&= \theop{f}{}{}\theop{g}{}{} \theop{H}{el}{}(\theel{}, \ther{}+ \thert{}, \theorig{} + \thet{}) \theop{g}{}{\dagger} \theop{f}{}{\dagger}  \nonumber \\
&= \theop{f}{}{} \theop{H}{el}{}(\theel{}, \ther{}+ \thert{}, \theorig{}) \theop{f}{}{\dagger}
\end{align}
The effect of $\theu{}{}$ is therefore to translate the nuclear coordinates, leaving the gauge origin and the electronic coordinates unchanged:
\begin{align}
\theu{}{}  (\theel{}, \ther{}, \theorig{},\thet{})
&\theop{H}{el}{}(\theel{}, \ther{}, \theorig{}) 
\theu{}{\dagger} (\theel{}, \ther{}, \theorig{},\thet{})
\nonumber \\
&=
\theop{H}{el}{}(\theel{}, \ther{} + \thert{}, \theorig{}) \label{UHUeq},
\end{align}
Within the Born--Oppenheimer approximation, the total electronic pseudomomentum operator therefore generates a translation of all nuclei in the system. Note that the phase factor cancels out such that eq.~\eqref{UHUeq} holds independently of our choice of $\theorig{}$ and $\eta(\ther{}, \theorig{}, \thet{})$. From now on, we will omit the arguments $\theel{}$ and $\theorig{}$ from operators and the argument $\theorig{}$ from wave functions to ease the reading of the equations.

Multiplying the electronic Schr\"odinger equation in eq.~\eqref{schrodinger} from the left by $\theu{}{} (\ther{}, \thet{})$ on both sides and recalling that $\theu{}{\dagger} (\ther{}, \thet{}) \theu{}{} (\ther{}, \thet{}) = 1$, we conclude from eq.~\eqref{UHUeq} that
\begin{align}
\label{FUF}
\thewf{el}{}{0}(\theel{}; \ther{} + \thert{}) &= \theu{}{} (\ther{}, \thet{}) \thewf{el}{}{0}(\theel{}; \ther{}) .
\end{align}
The operator $\theu{}{}  (\ther{}, \thet{})$ thus generates an eigenfunction of $\theop{H}{el}{}(\ther{} + \thert{})$ from an eigenfunction of $\theop{H}{el}{}(\ther{})$. Please keep in mind that $\theu{}{} (\ther{}, \thet{})$ gives us one particular solution $\thewf{el}{}{0}(\theel{}; \ther{} + \thert{})$ with a phase that depends on our choice of $\eta$. Consequently, the entire set of $\theu{}{}$'s generates the entire set of possible eigenfunctions, each differing by a phase.

\subsection{Translation of Total Nuclear Momentum Operator}
\label{secUPU}

In the present section, we consider the translation of the total nuclear canonical momentum operator, as obtained by the unitary transformation
\begin{align}
 \thepnucopu{\mathrm{tot}} (\ther{}, \thet{}) =
\theu{}{\dagger} &(\ther{},\thet{}) 
\thepnucop{\mathrm{tot}} 
\theu{}{} (\ther{},\thet{}). \label{PU}
\end{align}
Performing a Baker--Campbell--Hausdorff expansion, we obtain a series of commutators with $\theop{K}{}{} (\ther{}, \thet{})$ defined in eq.~\eqref{Uop}
\begin{align}
 \thepnucopu{\mathrm{tot}} (\ther{}, \thet{})
&= \thepnucop{\mathrm{tot}} + \left[\thepnucop{\mathrm{tot}}, \theop{K}{}{} (\ther{}, \thet{})\right] 
\nonumber \\
&\qquad+ \dfrac{1}{2} \left[\left[\thepnucop{\mathrm{tot}}, \theop{K}{}{} (\ther{}, \thet{})\right], \theop{K}{}{} (\ther{}, \thet{})\right] + \cdots \label{BCH}
\end{align}
Evaluating the commutators and noting that $\thepnucop{\mathrm{tot}} = - \thei \hbar \thenab{\mathrm{T}}$, we find
\begin{align}
\left[\thepnucop{\mathrm{tot}}, \theop{K}{}{} (\ther{}, \thet{})\right] &=  
- \thekelop{\mathrm{tot}} 
+ \thenab{\mathrm{T}} \eta(\ther{}, \thet{}), \\
\left[\left[\thepnucop{\mathrm{tot}}, \theop{K}{}{} (\ther{}, \thet{})\right], \theop{K}{}{} (\ther{}, \thet{})\right] &=  \dfrac{\theelcharge{}}{2} \then{el} \theb{} \times \thet{},
\end{align} 
where all third- and higher-order terms in eq.\,\eqref{BCH} vanish since $[\theb{} \times \thet{}, \theop{K}{}{} (\ther{}, \thet{})] = \mathbf 0$. It follows that 
\begin{align}
\thepnucopu{\mathrm{tot}}  (\ther{}, \thet{}) \!=\! - \thei \hbar \thenab{\mathrm{T}}
- \thekelop{\mathrm{tot}} + \thegamma{}(\ther{}, \thet{})
+ \dfrac{\theelcharge{}}{2} \then{el} \theb{} \times \thet{}, \label{UPU}
\end{align}
where $\thegamma{}$ is the derivative of $\eta$ with respect to $\thet{}$:
\begin{equation}
\thegamma{}(\ther{}, \thet{}) = \thenab{\mathrm{T}} \eta(\ther{}, \thet{}).
\end{equation}
We are now ready to consider the translation of the total Berry curvature and total Berry connections
in the next subsections.

\subsection{Total Berry Curvature for Exact Wave Functions}
\label{BCurvature}

Regarding the evaluation of the total Berry curvature, we note that it may be written in the manner
\begin{align}
\theom{\mathrm{tot}}{}{\ther{}} &= 
\dfrac{\thei}{\hbar}\Braket{
\thepnucop{\mathrm{tot}} \thewf{el}{}{0}(\ther{})| 
\thepnucop{\mathrm{tot}}^\mathrm{T} \thewf{el}{}{0}(\ther{})
}
\nonumber \\
& \quad 
- \dfrac{\thei}{\hbar}
\Braket{
\thepnucop{\mathrm{tot}} \thewf{el}{}{0}(\ther{})| 
\thepnucop{\mathrm{tot}}^\mathrm{T} \thewf{el}{}{0}(\ther{})
}^\mathrm{T}.
\label{pro_003a}
\end{align}
Since the nuclear canonical momentum is not Hermitian, we cannot use the turnover rule to simplify the equation.  Instead, we combine eqs.\,\eqref{FUF} and~\eqref{PU} to obtain
\begin{align}
\thepnucop{\mathrm{tot}}
\thewf{el}{}{0}(\theel{}; \ther{} + \thert{}) 
&= \thepnucop{\mathrm{tot}} \theu{}{} (\ther{}, \thet{}) \thewf{el}{}{0}(\theel{}; \ther{})  \nonumber \\
&= 
\theu{}{} (\ther{}, \thet{}) 
\hat {\mathbf P}^U_\text{tot}  (\ther{}, \thet{}) 
\thewf{el}{}{0}(\theel{}; \ther{}), 
\end{align}
which at $\thet{} = \mathbf 0$ according to eq.~\eqref{UPU} reduces to
\begin{equation}
\thepnucop{\mathrm{tot}} \thewf{el}{}{0}(\theel{}; \ther{})
= \left[ \thegamma{}(\ther{}) - \thekelop{\mathrm{tot}} \right]
\thewf{el}{}{0}(\theel{}; \ther{}), 
\label{Udef3}
\end{equation}
in the notation $\thegamma{}(\ther{})= \thegamma{}(\ther{},\mathbf 0)$.
Inserting this expression into eq.\,\eqref{pro_003a} and using the turnover rule, we obtain
\begin{align}
\theomtotel{\indx}{\indy}{\ther{}}
&=\frac{\thei}{\hbar} 
\Braket{ \thewf{el}{}{0}(\ther{})|
\left[
\thekeldir{\mathrm{tot}}{\indx} \!-\! \thegammael{}{\indx} (\ther{}), \thekeldir{\mathrm{tot}}{\indy} \!-\! \thegammael{}{\indy} (\ther{})
\right]
| \thewf{el}{}{0}(\ther{})}
\nonumber \\
&=\frac{\thei}{\hbar}
\Braket{ \thewf{el}{}{0}(\ther{})|
\left[
\thekeldir{\mathrm{tot}}{\indx}, \thekeldir{\mathrm{tot}}{\indy}
\right]
| \thewf{el}{}{0}(\ther{})}
\nonumber \\
&= -\theelcharge{}\then{el} \epsilon_{\indx\indy\indz}  \theb{\indz} 
,
\label{tra_014c}
\end{align}
where, in the last step, we have used the commutator in eq.\,\eqref{comkk}. Since eq.~\eqref{tra_014c}  gives eq.~\eqref{mtsr} in matrix notation, we conclude that the magnetic-translational sum rule holds for an exact wave function. This result confirms the conclusion of Yin and Mead's more heuristic arguments.\cite{Yin1992}

\subsection{Total Berry Connection for Exact Wave Functions}
\label{BConnection}

In the present section, we study the translational behavior of the total Berry connection $\thechi{\mathrm{tot}}{\ther{},\theorig{}}$,
\begin{align}
\thechi{\mathrm{tot}}{\ther{}} &=
\sum_{\indi=1}^{\then{nuc}} \thechi{\indi}{\ther{}} = 
\Braket{ \thewf{el}{}{0}(\ther{})| 
\thepnucop{\mathrm{tot}} \thewf{el}{}{0}(\ther{}) }. \label{ham_007}
\end{align}
Since the Berry connection plays the role of a vector potential and since the vector potential in eq.~\eqref{Avec} translates as $\thea{\thet{}}{\theorig{}} - \thea{\mathbf{0}}{\theorig{}} = \frac{1}{2} \theb{} \times \thet{}$, we expect the Berry connection to behave in a similar manner upon translation of the nuclear coordinates. Expressing the expectation values in terms of the transformed nuclear momentum operator in eq.~\eqref{PU}, we obtain
\begin{align}
\Delta \thechi{\mathrm{tot}}{\ther{},\thet{}} &= 
\thechi{\mathrm{tot}}{\ther{} + \thert{}} - \thechi{\mathrm{tot}}{\ther{}} \label{chiRT} 
\nonumber \\ &= 
\Braket{ \thewf{el}{}{0}(\ther{}) | 
\left[ \thepnucopu{\mathrm{tot}} (\ther{}, \thet{}) \!-\! 
\thepnucopu{\mathrm{tot}} (\ther{}, \mathbf 0)  \right] 
\thewf{el}{}{0}(\ther{}) } .
\end{align}
Use of eq.~\eqref{UPU} yields
\begin{equation}
\Delta \thechi{\mathrm{tot}}{\ther{},\thet{}} 
= \dfrac{\theelcharge{}}{2} \then{el}  (\theb{} \times \thet{}) + \Delta \thegamma{}(\ther{}, \thet{}), \label{pro_002}
\end{equation}
where $ \Delta \thegamma{}(\ther{}, \thet{})$ is the change in the phase factor:
\begin{equation}
 \Delta \thegamma{}(\ther{}, \thet{})  = 
 \thegamma{}(\ther{}, \thet{}) - \thegamma{}(\ther{}, \mathbf 0) .
\end{equation}
The first term in eq.\,\eqref{pro_002} has the form we would expect for the translation of a vector potential (multiplied by the total electronic charge $-\then{el} e$). The second term vanishes when the gauge function is chosen to be translationally invariant.

Finally, as an alternative to the derivation given in Section\;\ref{BCurvature}, we may obtain the total Berry curvature by differentiation of the total Berry connection, using eq.~\eqref{ham_006a} in combination with eq.~\eqref{chiRR} and summing over all pairs of nuclei:
\begin{align}
\theom{\mathrm{tot}}{}{\ther{}+\thert{}} &=  
\left[\thenab{\mathrm{T}} \thechit{\mathrm{tot}}{\ther{} + \thert{}} \right]^\text T
\nonumber \\
&\qquad-  \thenab{\mathrm{T}} \thechit{\mathrm{tot}}{\ther{} + \thert{}}.
\label{pro_004x}
\end{align}
Differentiation of the expression given in eq.\,\eqref{pro_002} gives
\begin{align}
\thenab{\mathrm{T}}  &\thechit{\mathrm{tot}}{\ther{} + \thert{}} =
\thenab{\mathrm{T}} \Delta \thechit{\mathrm{tot}}{\ther{},\thet{}} \nonumber \\
&= \dfrac{\theelcharge{}}{2} \then{el}  \thenab{\mathrm{T}} (\theb{} \times \thet{})^{\text T} 
+ \thenab{\mathrm{T}}  \thegamma{}^{\text T}(\ther{}, \thet{}).
\end{align}
Noting next that
\begin{align}
\thenab{\mathrm{T}} \left(\theb{}\times \thet{} \right)^\mathrm T 
&= -\thebt{}, \\
\left[ \thenab{\mathrm{T}} \thegamma{}^\mathrm T(\ther{},  \thet{})\right]^\mathrm T &= \thenab{\mathrm{T}} \thenab{\mathrm{T}}^\mathrm T  \, \eta (\ther{}, \thet{}),
\end{align}
we obtain
\begin{align}
\thenab{\mathrm{T}} \thechit{\mathrm{tot}}{\ther{} + \thert{}}&= 
-\frac{\theelcharge{}}{2} \then{el} \thebt{} + \thenab{\mathrm{T}} \thenab{\mathrm{T}}^\mathrm T  \, \eta (\ther{}, \thet{}).
\end{align}
Since $\thebt{}$ is antisymmetric while the contribution from the gauge function is symmetric, it follows that
\begin{align}
\theom{\mathrm{tot}}{}{\ther{}+\thert{}} 
=  \theelcharge{}\then{el} \thebt{}
\label{pro_004}
\end{align}
in agreement with the magnetic-translational sum rule [see eq.~\eqref{mtsr}].

\subsection{Berry Curvature in Exact Hartree--Fock theory}
\label{HartreeFock}

In exact HF theory, the MOs ($\{\themo{\indmo}{}{} (\theel{};\ther{}, \theorig{})\}$) may be taken to satisfy the canonical Fock equations
\begin{align}
\Braket{\themo{\indmo}{}{}(\ther{}) | \themo{\indmotwo}{}{} (\ther{})} &= \delta_{ij}, \\
\theop{F}{\varphi}{}(\ther{}) \themo{\indmo}{}{} (\theel{};\ther{}) &= \varepsilon_i (\ther{}) \themo{\indmo}{}{} (\theel{};\ther{})
\label{Feq}
\end{align}
We note that $\theel{}$ now refers to a single electronic coordinate and that the Fock operator $\theop{F}{\varphi}{}(\theel{},\ther{}, \theorig{})$ depends on the MOs but transforms in the same manner as the many-electron Hamiltonian  $\theop{H}{el}{}(\theel{}, \ther{})$ in eq.~\eqref{Htra}, having the same structure of the kinetic operator. In accordance with Section~II E, we can thus set up a unitary operator $\theu{\indmo}{} (\theel{}, \ther{}, \theorig{},\thet{})$ as
\begin{align}
\theu{\indmo}{} (\ther{}, \thet{}) &= 
\exp\left\{ - \dfrac{\thei}{\hbar} \thet{} \cdot \thekelop{} + \dfrac{\thei}{\hbar} \eta_\indmo(\ther{}, \thet{}) \right\}
\end{align}
that induces a translation by $\thet{}$ of the nuclei in the Fock operator and the $i$th MO:
\begin{align}
\theop{F}{\varphi}{}(\ther{} + \thert{}) &= 
\theu{\indmo}{}(\ther{}, \thet{})\theop{F}{\varphi}{}(\ther{}) \theu{\indmo}{\dagger}(\ther{}, \thet{})
\label{UHFUeq} \\
\label{HFUF}
\themo{\indmo}{}{}(\theel{};\ther{} + \thert{}) &=
\theu{\indmo}{} (\ther{}, \thet{}) \themo{\indmo}{}{}(\theel{};\ther{})
\end{align}
Note that $\thekelop{} (\theel{}, \theorig{})$ now acts on a single electron and that the use of a different gauge function $\eta_\indmo (\ther{}, \theorig{}, \thet{})$ for each MO leaves eq.~\eqref{UHFUeq} unchanged. Proceeding as for the exact many-electron wave function, we find that
\begin{align}
\thepnucop{\mathrm{tot}} \themo{\indmo}{}{}(\theel{}; \ther{}) &=
\left[ \thegamma{\indmo} (\ther{}) - \thekelop{} \right]
\themo{\indmo}{}{}(\theel{}; \ther{}) \label{ldn_009d} \\
\thegamma{\indmo} (\ther{}) &= \thenab{\mathrm{T}}  \eta_\indmo (\ther{},  \thet{}) |_{\thet{} = \mathbf{0}}
\end{align}
corresponding to eq.~\eqref{Udef3} for the exact wave function.

We now consider the evaluation of the total Berry curvature in HF theory. Using eq.~(55) of ref.~\onlinecite{Culpitt2022} and taking into account that the final two terms therein are real-valued, the Cartesian components of the total Berry curvature can be written in terms of MOs as
\begin{align}
\theomtotel{\indx}{\indy}{\ther{}}
&= 
- \dfrac{2}{\hbar} \Im \Bigg \{
\sum \limits_{\indmo=1}^{\then{occ}}
\Braket{
\thepnucdir{\mathrm{tot}}{\indx} \themo{\indmo}{}{} (\ther{}) 
|
\thepnucdir{\mathrm{tot}}{\indy} \themo{\indmo}{}{} (\ther{}) 
}
\Bigg \}.
\label{ldn_009bx}
\end{align}
Combining eq.\,\eqref{ldn_009bx} with eq.\,\eqref{ldn_009d}, we can make use of the turnover rule
\begin{align}
\theomtotel{\indx}{\indy}{\ther{}}
&=  \dfrac{\thei{}}{\hbar} 
\sum \limits_{\indmo=1}^{\then{occ}}
\Braket{
\themo{\indmo}{}{} (\ther{}) 
|
\left[
\thekeldir{}{\indx}, \thekeldir{}{\indy}
\right]
|
\themo{\indmo}{}{} (\ther{}) 
}
\nonumber \\
&= -\theelcharge{}\then{el} \epsilon_{\indx\indy\indz}  \theb{\indz} 
,
\label{ldn_009by}
\end{align}
to conclude that the total Berry curvature in the HF basis-set limit correctly reproduces the magnetic-translational sum rule. However, this relation may not hold for in a finite orbital basis since the approximate MOs do not necessarily transform according to eq.~\eqref{HFUF}.

\subsection{Hartree--Fock Berry Curvature with London orbitals}
\label{HartreeFockLDN}

We next consider HF wave functions with MOs expanded in a finite set of London atomic orbitals  $\theldn{\indmu}{}{}(\theel{}; \ther{\indmu})$  in the manner
\begin{equation}
\themo{\indp}{}{} (\theel{};\ther{}) = 
\sum \limits_{\indmu = 1}^{\then{bas}} \thecoef{\indmu}{\indp}{0}(\ther{}) \theldn{\indmu}{}{}(\theel{}; \ther{\indmu}), \label{MOs}
\end{equation}
where each London  orbital $\theldn{\indmu}{}{}(\theel{}; \ther{\indmu}, \theorig{}) $ is a standard Gaussian atomic orbital $\thegsn{\indmu}{}{}(\theel{} - \ther{\indmu})$ multiplied by a field-dependent phase factor:
\begin{equation}
\theldn{\indmu}{}{}(\theel{}; \ther{\indmu}) = \exp
\left\{
-\dfrac{\thei \theelcharge}{2 \hbar} \theb{} \times (\ther{\indmu} - \theorig{}) \cdot \theel{} 
\right\}
\thegsn{\indmu}{}{}(\theel{} - \ther{\indmu}).
\end{equation}
The expansion coefficients $\thecoef{\indmu}{\indp}{0}(\ther{})$ are at each $\ther{}$ chosen such that the MOs are orthonormal and the Fock matrix diagonal:
\begin{align}
\Braket{\themo{\indp}{}{} (\ther{}) | \themo{\indq}{}{} (\ther{})} &= \delta_{\indp\indq}, \label{ldnfock}\\
\Braket{ \themo{\indp}{}{}(\ther{}) | \theop{F}{\varphi}{}(\ther{}) | \themo{\indq}{}{} (\ther{})} 
&= \varepsilon_p (\ther{}) \,\delta_{\indp\indq}. \label{ldnfock2}
\end{align}
The Fock operator transforms in the same manner as in exact HF theory upon a nuclear translation; see eq.~\eqref{UHFUeq}. We shall now see that it is possible to choose the dependence of the optimal MOs on $\ther{}$ such that they transform in the same manner as the MOs in exact HF theory upon a translation of all nuclei [see eq.~\eqref{HFUF}].

Carrying out some straightforward algebra, we find that the London atomic orbitals transform as the exact wave function but with an additional global phase factor that depends on its center:
\begin{align}
\theu{\indp}{}(&\ther{}, \thet{}) \theldn{\indmu}{}{}(\theel{}; \ther{\indmu})
\nonumber \\
&=   
\exp\left\{
\dfrac{\thei \theelcharge}{2 \hbar} 
\left[\theb{} \times (\ther{\indmu} - \theorig{}) 
\right]\cdot \thet{} \right\}
\theldn{\indmu}{}{}(\theel{}; \ther{\indmu}+ \thet{})
\end{align}
Since we are allowed to multiply the coefficients with this phase factor leaving all observables unchanged
\begin{align}
\thecoef{\indmu}{\indp}{0}(&\ther{} + \thert{}) = \exp\left\{
\dfrac{\thei \theelcharge}{2 \hbar} 
\left[\theb{} \times (\ther{\indmu} - \theorig{}) 
\right]\cdot \thet{} \right\}
\thecoef{\indmu}{\indp}{0}(\ther{}),  \label{MOT}
\end{align}
we find that the approximate MOs in eq.~\eqref{MOs} transform in the same manner as the exact MOs upon a nuclear translation:
\begin{align}
\label{HFUF0}
\theu{\indp}{} (\ther{},\thet{}) \themo{\indp}{}{} (\theel{};\ther{})
&= 
\sum \limits_{\indmu = 1}^{\then{bas}} 
\thecoef{\indmu}{\indp}{0}(\ther{}) \theu{\indp}{} (\ther{},\thet{}) \theldn{\indmu}{}{}(\theel{}; \ther{\indmu})
\nonumber \\
&=\sum \limits_{\indmu = 1}^{\then{bas}} 
\thecoef{\indmu}{\indp}{0}(\ther{} + \thert{}) \theldn{\indmu}{}{}(\theel{}; \ther{\indmu} + \thet{})
\nonumber \\
&=
\themo{\indp}{}{}(\theel{};\ther{} + \thert{} ) .
\end{align}
Consequently, we can use the same argument as in the previous subsection to demonstrate that the magnetic-translational sum rule is fulfilled when the MOs are expanded in a finite set of London atomic orbitals. This has also been observed in molecular simulations in strong magnetic fields.\cite{Ceresoli2007,Peters2021,Monzel2022}

\subsection{Mulliken Approximations to the Berry Curvature}
\label{Mulliken}

Equation~\eqref{ldn_009by} gives us the opportunity to derive (reasonable) approximations to the Berry curvature. This is particularly important for molecular dynamics simulations, where the calculation of the Berry curvature adds appreciable computational expense. We start by assuming that every component of the Berry curvature (not only their sum) can be written in terms of nuclear fragments of the total electronic pseudomomentum ($\thekelop{\indi}$): 
\begin{align}
\theom{\indi}{\indj}{\ther{}} &\approx 
\Re \Bigg\{
\dfrac{\thei}{\hbar} 
\sum \limits_{\indmo=1}^{\then{occ}}
\Braket{
\themo{\indmo}{0} (\ther{})
|
\thekelop{\indi}
\thekelop{\indj}^\mathrm{T}
-
\left[
\thekelop{\indj}
\thekelop{\indi}^\mathrm{T}
\right]^\mathrm{T}
|
\themo{\indmo}{0} (\ther{})
}
\Bigg\}
\label{mul_000}
\end{align}
By enforcing that the Berry curvature is real-valued, we also allow for $\thekelop{\indi}$ that are not Hermitian operators. In order to ensure the correct properties, the operators $\thekelop{\indi}$ must reproduce the commutator of the electronic pseudomomentum
\begin{align}
\sum \limits_{\indi,\indj=1}^{\then{nuc}}
\sum \limits_{\indmo=1}^{\then{occ}}
\Braket{
\themo{\indmo}{0} (\ther{}) |
\left[
\thekeldir{\indi}{\indx} , \thekeldir{\indj}{\indy}
\right] |
\themo{\indmo}{0} (\ther{})
}
&= \thei{} \hbar \varepsilon_{\indx\indy\indz} \theelcharge{} \then{el}\theb{\indz}
\label{mul_001}
\end{align}
The remaining question is now how to separate $\thekelop{}$ into the contributions from the different nuclei. 

In our first ansatz, we assume that the electrons are tightly bound to the nuclei according to the Mulliken partitioning of the electron density. Consequently, we can replace the electronic coordinate $\theel{}$ with a nuclear coordinate and introduce the Mulliken projector $\themulop{\indi}$\cite{Carbo-Dorca2004} that we treat as independent of the nuclear and electronic coordinates:
\begin{align}
\thekelop{\indi}^\mathrm{M1}  &= 
- \thei \hbar \dfrac{\partial}{\partial \ther{\indi}}  - \dfrac{\theelcharge{}}{2} \theb{} \times [\ther{\indi} + \theorig] \themulop{\indi} \label{mul_002} \\
\themulop{\indi} &= \sum \limits_{\mu \in \indi} \sum \limits_{\indnu}^{\then{bas}}
\ket{\theldn{\indmu}{}{}(\ther{})}
\Braket{\theldn{\indmu}{}{}(\ther{})|\theldn{\indnu}{}{}(\ther{})}^{-1}
\bra{\theldn{\indnu}{}{}(\ther{})} 
\label{mul_002b}
\end{align}
Inserting this first Mulliken approximation (M1) into eq.~\eqref{mul_000}, we see that the resulting Berry curvature $\theoms{\indi}{\indj}{\ther{}}{M1}$ depends only on the electronic part of the Mulliken charges\cite{Mulliken1955} 
\begin{align} 
\theqm{\indi}{}{\ther{}} 
 = - \dfrac{1}{2}
&\sum \limits_{\indmo=1}^{\then{occ}}
\big[
\Braket{
\themo{\indmo}{0} (\ther{})
|
\themulop{\indi}\themo{\indmo}{0} (\ther{})
}
\nonumber \\
&\qquad+
\Braket{
\themulop{\indi}\themo{\indmo}{0}  (\ther{})
|
\themo{\indmo}{0} (\ther{})
}
\big] 
\label{mul_003}
\end{align}
as well as the magnetic field:
\begin{align} 
\theoms{\indi}{\indj}{\ther{}}{M1} &= - \delta_{\indi\indj} \theelcharge{} \theqm{\indi}{}{\ther{}} \thebt{}
\label{mul_004}
\end{align}
Since the Berry force of $\indi$ depends only on its charge and velocity
\begin{align}
\thef{\indi}{M1} (\ther{},\therdot{}) &= \sum \limits_{\indj=1}^{\then{nuc}} \theoms{\indi}{\indj}{\ther{}}{M1} \therdot{\indj} = - \theelcharge{} \theqm{\indi}{}{\ther{}} \thebt{} \therdot{\indi} ,
\label{mul_005}
\end{align}
the use of M1 is equivalent to simulating the system with an effective nuclear charge calculated from the Mulliken charge at a given molecular geometry:
\begin{align}
\thef{\indi}{L} (\therdot{}) + \thef{\indi}{M1} (\ther{},\therdot{})  &= - \theelcharge{} [\thez{\indi} + \theqm{\indi}{}{\ther{}} ] \thebt{} \therdot{\indi}
\label{mul_006}
\end{align}
Thus, we assume that each nucleus is screened by the amount of electrons it has been assigned to according to the Mulliken scheme. 

In the second Mulliken scheme (M2), we also use the Mulliken partitioning but keep the original electronic-coordinate dependence of the total electronic pseudomomentum:
\begin{align}
\thekelop{\indi}^\mathrm{M2}  &= 
- \thei \hbar \themulop{\indi}^* \dfrac{\partial}{\partial \theel{}}   - \dfrac{\theelcharge{}}{2} \theb{} \times [\theel{} + \theorig] \themulop{\indi} 
\label{mul_007}
\end{align}
As a result of this, the Berry curvature now contains the Mulliken overlap populations [$\theqm{\indi}{\indj}{\ther{}}$]:
\begin{align} 
\theoms{\indi}{\indj}{\ther{}}{M2} &= 
-  \theelcharge{} \theqm{\indi}{\indj}{\ther{}} \thebt{} 
\label{mul_008} \\
\theqm{\indi}{\indj}{\ther{}} 
 = - \dfrac{1}{2}
&\sum \limits_{\indmo=1}^{\then{occ}}
\big[
\Braket{
\themulop{\indj}\themo{\indmo}{0} (\ther{})
|
\themulop{\indi}\themo{\indmo}{0} (\ther{})
}
\nonumber \\
&\qquad+
\Braket{
\themulop{\indi}\themo{\indmo}{0} (\ther{})
|
\themulop{\indj}\themo{\indmo}{0} (\ther{})
}
\big] 
\label{mul_009}
\end{align}
From the resulting Berry forces,
\begin{align}
\thef{\indi}{M2} (\ther{},\therdot{}) &= - \theelcharge{} \sum \limits_{\indj=1}^{\then{nuc}} \theqm{\indi}{\indj}{\ther{}} \thebt{} \therdot{\indj} ,
\label{mul_010}
\end{align}
we see that the second approximation introduces a coupling between the motion of the different nuclei via the electrons, since the Berry force of $\indi$ depends on the motion of nucleus $\indj$. In this context, $\theqm{\indi}{\indj}{\ther{}}$ can be interpreted as the amount of electrons of nucleus $\indj$ that screen the nucleus $\indi$ according to the Mulliken scheme. Here, the atomic Mulliken charge corresponds to the effective nuclear charge that arises when the entire molecular system moves with a constant velocity in a magnetic field:
\begin{align}
\thef{\indi}{L} (\therdot{\mathrm{T}}) + \thef{\indi}{M2} (\ther{},\therdot{\mathrm{T}}) &= - \bigg[\thez{\indi} +  \sum \limits_{\indj=1}^{\then{nuc}} \theqm{\indi}{\indj}{\ther{}} \bigg]  \thebt{} \thetdot{} 
\label{mul_011}
\end{align}

In our final approximation, we represent the electronic density by Mulliken dipole moments [$\thedip{\indi}{}{\ther{}}$]:
\begin{align}
\thekelop{\indi}^\mathrm{M3}  &= 
- \thei \hbar \dfrac{\partial}{\partial \ther{\indi}}  
+ \dfrac{\theelcharge{}}{2 \then{el}} \theb{} \times \thedip{\indi}{}{\ther{}} 
\label{mul_012}
\\
\thedip{\indi}{}{\ther{}} = - \dfrac{1}{2}
&\sum \limits_{\indmo=1}^{\then{occ}}
\big[
\Braket{
\themo{\indmo}{0} (\ther{})
| \theel{}|
\themulop{\indi}\themo{\indmo}{0} (\ther{})
}
\nonumber \\
&\qquad+
\Braket{
\themulop{\indi}\themo{\indmo}{0} (\ther{})
| \theel{}|
\themo{\indmo}{0} (\ther{})
}
\big] 
\label{mul_015}
\end{align}
Consequently, the resulting Berry curvature in the third Mulliken approximation (M3) contains a polarization tensor [$\thefluc{\indi}{\indj}{\ther{}}{0}$], which is calculated as the derivative of the atomic dipole moment with respect to a nuclear coordinate:
\begin{align}
\theoms{\indi}{\indj}{\ther{}}{M3} &= - \dfrac{\theelcharge{}}{2} \left[
\thebt{} \thefluc{\indi}{\indj}{\ther{}}{0} + \thefluc{\indj}{\indi}{\ther{}}{1} \thebt{} \right]
\label{mul_013}
\\
\thefluc{\indi}{\indj}{\ther{}}{0} &= \dfrac{\partial \thedip{\indi}{}{\ther{}} }{\partial \ther{\indj}}
\label{mul_014}
\end{align}
Note that this Berry curvature still satisfies the magnetic-translational sum rule, since the atomic dipole moments sum up to the total electronic dipole moment [$\thedip{\mathrm{tot}}{}{\ther{}}$]:
\begin{align}
\sum \limits_{\indi,\indj=1}^{\then{nuc}} \thefluc{\indi}{\indj}{\ther{}}{0} &= 
\sum \limits_{\indj=1}^{\then{nuc}} \dfrac{\partial \thedip{\mathrm{tot}}{}{\ther{}} }{\partial \ther{\indj}} = -\theone{} \then{el}
\label{mul_016}
\end{align}
Equation~\eqref{mul_013} implies that the screening of nucleus $\indi$ by the electrons of $\indj$ can be described via the change of both atomic dipole moments (calculated according to the Mulliken scheme) with respect to the nuclear coordinates. 

We conclude this subsection by briefly discussing the computational costs of our approximations at the HF level of theory. The calculations of the M1 and M2 Berry curvatures do not depend on derivatives of orbitals with respect to nuclear coordinates. Consequently, they can be obtained with no additional cost subsequent to an energy calculation. This makes their calculation significantly faster than the calculation of the M3 and exact Berry curvatures, which both require solving the coupled-perturbed HF equations or carrying out numerical differentiation. Although the M3 model does not reduce the computation time, it may give insight into the interpretation of the exact Berry curvature.

\section{Computational Details}

All calculations were performed with the {\sc London} program package.\cite{London} We employed the HF method using a magnetic field of $10^{-3}B_0$. If not stated otherwise, we used the London orbital version of the contracted Gaussian cc-pVDZ basis set\cite{Dunning1989}, denoted by l-cc-pVDZ. The molecular geometries of H$_2$, LiH, BH$_3$, CH$_4$, NH$_3$, H$_2$O, FH, and CH$_3$OH were optimized at the HF/l-cc-pVDZ/$|\theb{}| = 10^{-3}B_0$ level of theory prior to calculating the exact Berry curvature \textit{via} finite differences as presented in ref.~\onlinecite{Culpitt2021} as well as the approximate Berry curvatures M1 [eq.~\eqref{mul_004}], M2 [eq.~\eqref{mul_008}], and M3 [eq.~\eqref{mul_013}]. For H$_2$, LiH, CH$_4$, H$_2$O, FH, and CH$_3$OH, we used a local minimum of the geometry with the principal axes perpendicular to the magnetic field, while the axes of BH$_3$ and NH$_3$ were parallel to the field. In all calculations, we place the center of mass as well as the gauge origin ($\theorig{}$) at $(0,0,0)$. 

We determine three Euclidean error measures: $\epsilon_\mathrm{com}$ as the error in the center-of-mass motion (resulting from violation of the magnetic-translational sum rule) during dynamics [see eq.~\eqref{mtsr}],
\begin{align}
\epsilon_\mathrm{com} = \big|\big| 
\theom{\mathrm{tot}}{}{\ther{}}  - \theelcharge{}\then{el} \thebt{} \big|\big|_2 ,
\label{error}
\end{align}
$\epsilon_\mathrm{tot}^\mathrm{MX}$ as the error in \emph{all} elements of the approximate Berry curvature MX = M1, M2, and M3,
\begin{align}
\epsilon_\mathrm{tot}^\mathrm{MX} = 
\big|\big| 
\theoms{}{}{\ther{}}{MX}  - \theom{}{}{\ther{}} \big|\big|_2 ,
\label{error2}
\end{align}
and $\epsilon_{\indi\indj}^\mathrm{MX}$ as the error in the elements within a single $\indi{}\indj{}$ block of the approximate Berry curvature
\begin{align}
\epsilon_{\indi\indj}^\mathrm{MX} = 
\big|\big| 
\theoms{\indi}{\indj}{\ther{}}{MX}  - \theom{\indi}{\indj}{\ther{}} \big|\big|_2 .
\label{error3}
\end{align}
For the latter, we calculate the average as well as the maximum values:
\begin{align}
\epsilon_\mathrm{avg}^\mathrm{MX} &= \dfrac{1}{\then{nuc}^2}
\sum \limits_{\indi,\indj=1}^{\then{nuc}} 
\epsilon_{\indi\indj}^\mathrm{MX}, \label{error4} \\ 
\epsilon_\mathrm{max}^\mathrm{MX}  &= \mathrm{max} \{
\epsilon_{00}^\mathrm{MX}, \epsilon_{01}^\mathrm{MX}, ..., \epsilon_{\then{nuc}\then{nuc}}^\mathrm{MX} \} .
\label{error5}
\end{align}
To obtain a measure of screening error per electron, we divide all errors by the magnetic field strength times the number of electrons ($|\theb{}| \then{el}$). The scans of H$_2$ and LiH were carried out in a perpendicular field orientation, with bond distances in the range 0.2--10.0\,bohr for H$_2$ and $2.0--10.0$\,bohr for LiH, in steps of 0.1\,bohr.

\section{Results and Discussion}

\begin{figure*}
\centering
\begin{tabular}{ll}
(a) & (b) \\
\includegraphics[width=0.48\textwidth]{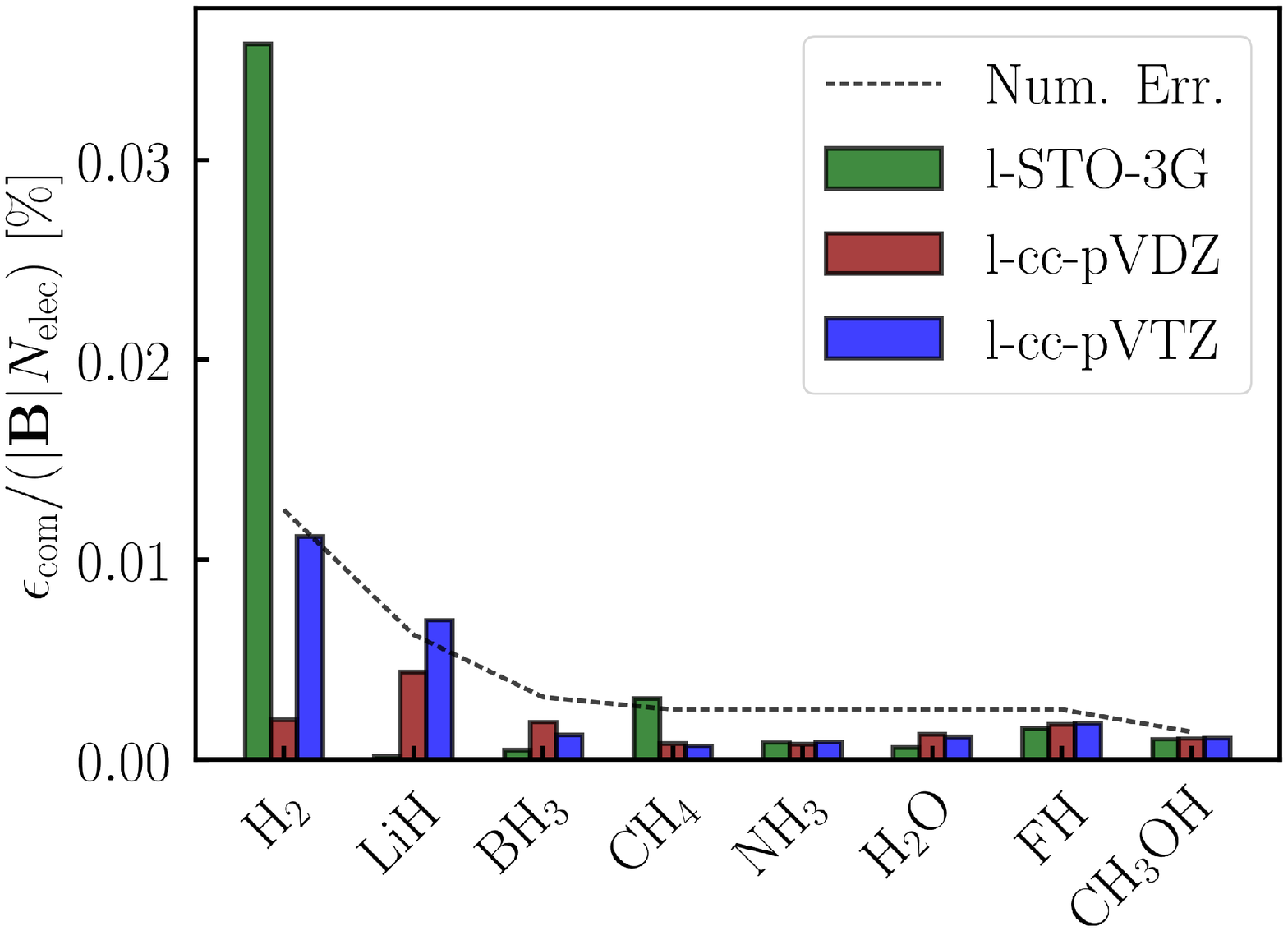} &
\includegraphics[width=0.48\textwidth]{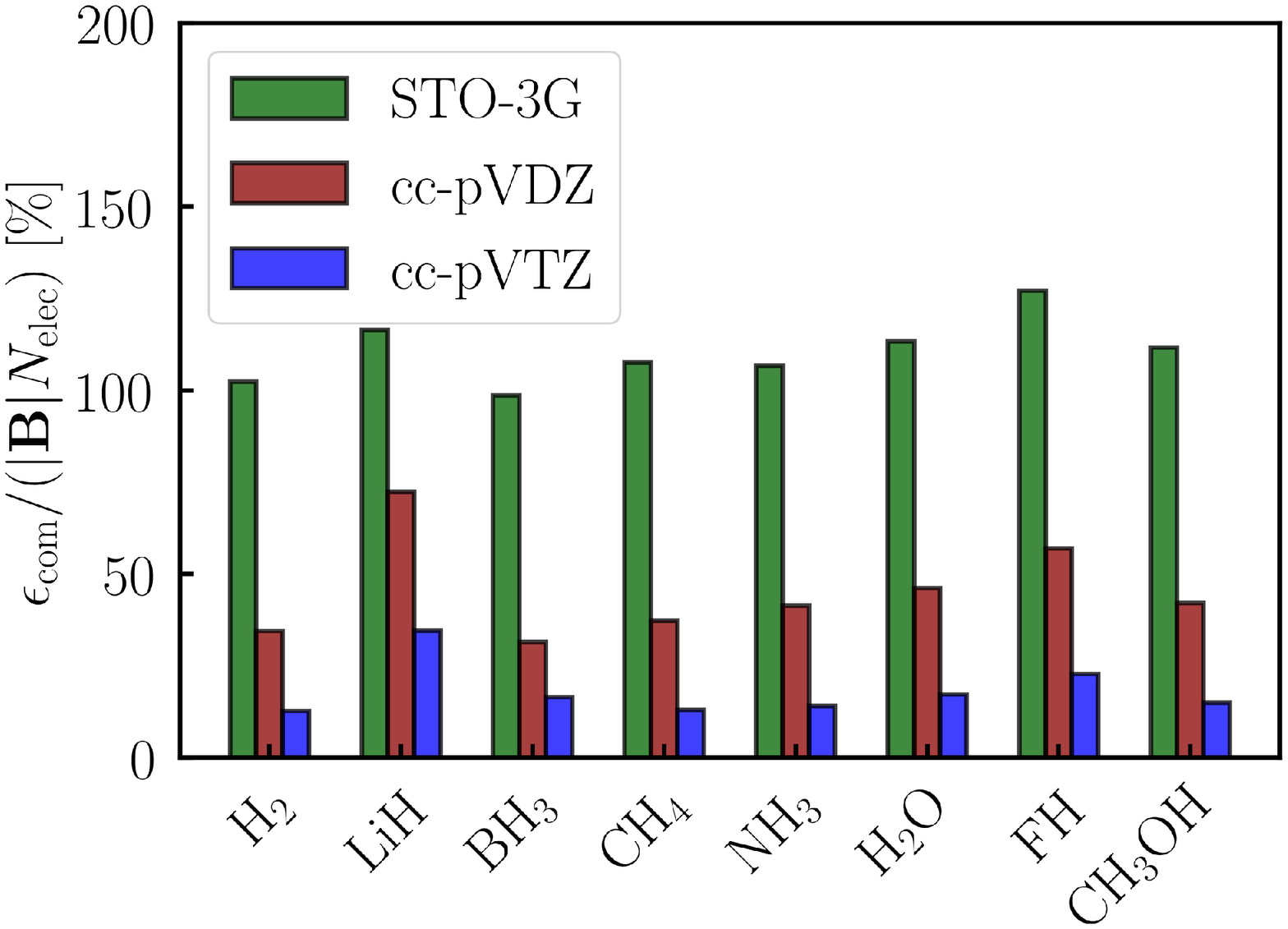} 
\end{tabular}
\caption{Error in the center-of-mass motion [see eq.~\eqref{error}] of the Berry curvature calculated for different molecules using London orbital basis sets (a) and regular Gaussian basis sets (b). All calculations were performed at the HF level of theory with $|\mathbf{B}| = 10^{-3}\,B_0$. We divide the individual error by the magnetic field strength and the number of electrons to get an estimate for the screening error per electron. The expected numerical error of the finite difference calculation [$2.5 \times 10^{-7}\,$a.u.] is indicated by the doted line in (a). Please note the different scale of the y-axis in (a) and (b).}
\label{fig1}
\end{figure*}

In the theory section, we showed that the correct (overall) center-of-mass motion during dynamics is closely connected to the magnetic-translational sum rule [see eq.~\eqref{mtsr}], which holds when London orbitals are used as basis sets. To illustrate this, we calculated the error according to eq.~\eqref{error} for a series of small molecules and basis sets with and without London phase factors [see Fig.~\ref{fig1}]. The errors obtained with London orbitals [see Fig.~\ref{fig1}(a)] are close to the error of the finite difference procedure used to determine the total Berry curvature [about $2.5 \times 10^{-7}\,$a.u.]. They can thus be regarded as numerical noise that is \emph{independent} of the size of the basis set. In contrast to this, the calculations without London orbitals [see Fig.~\ref{fig1}(b)] exhibit large errors that decrease with increasing basis-set size. While this behavior indicates that regular Gaussian basis sets reproduce the correct physics in the basis-set limit, they cannot be used in practice since the relatively large triple-zeta basis set cc-pVTZ still leads to errors of about 20\%, in agreement with our results for atoms in ref.~\onlinecite{Culpitt2021}. At this point we want to stress that the total Berry curvature calculated without London orbitals strongly depends on the nuclear coordinates and the gauge origin. Both dependencies vanish when using either the exact wave function or London orbitals.

\begin{figure*}
\centering
\begin{tabular}{ll}
(a) & (b) \\
\includegraphics[width=0.48\textwidth]{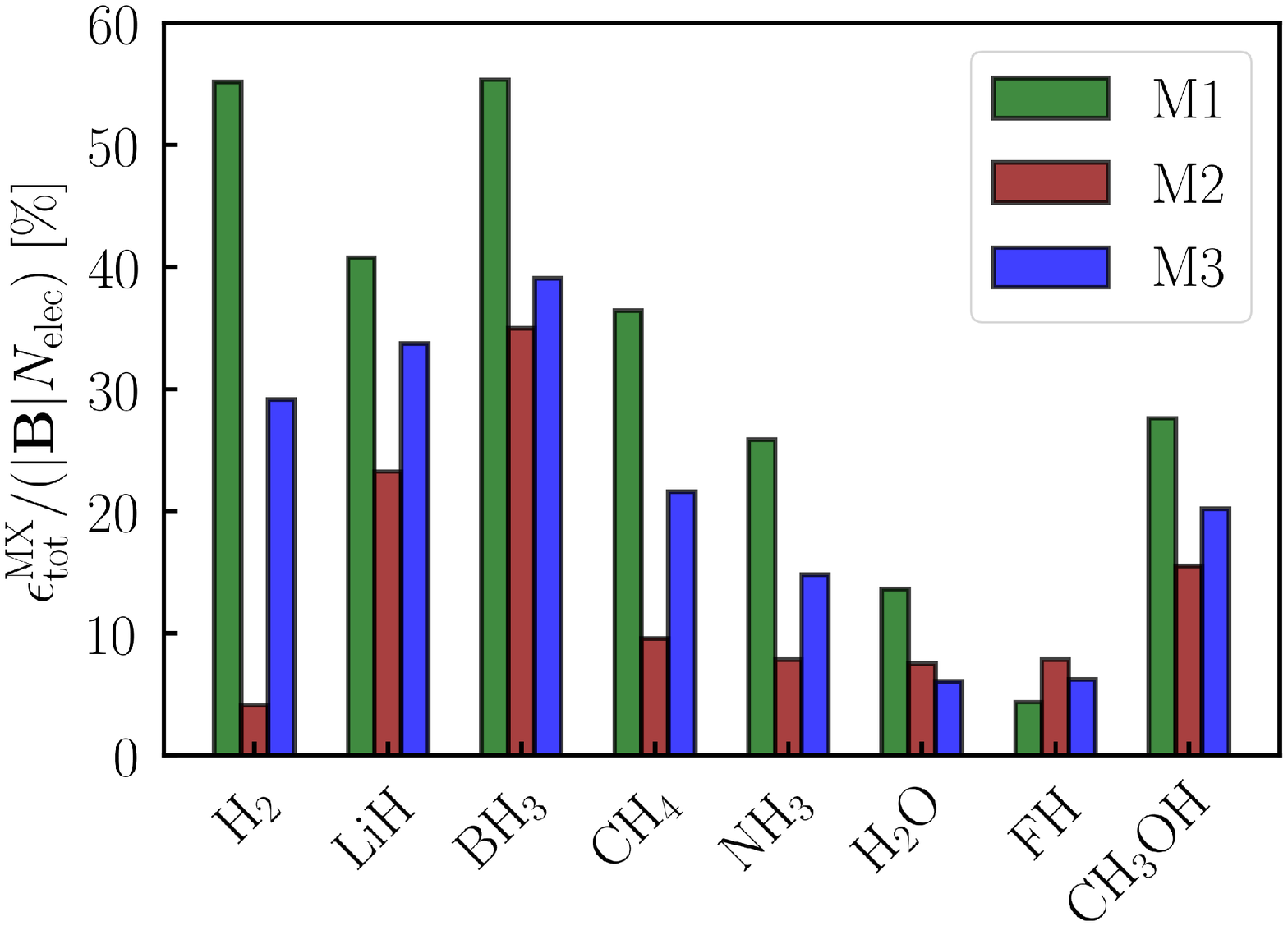} &
\includegraphics[width=0.48\textwidth]{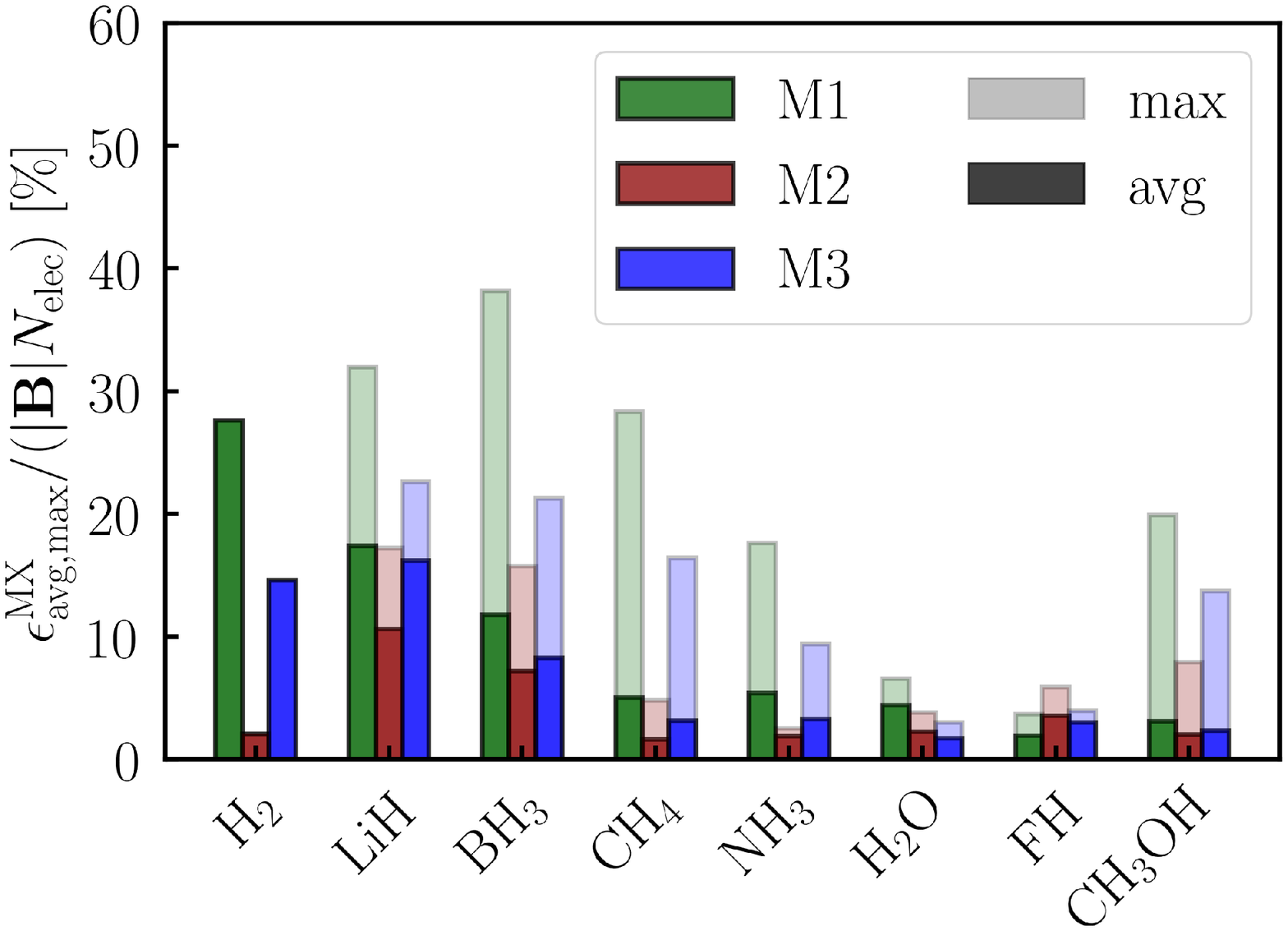} 
\end{tabular}
\caption{Overall [a, see eq.~\eqref{error2}] and average/maximum [b, see eqs.~\eqref{error4} and \eqref{error5}] error of the approximate Berry curvatures M1, M2, and M3 calculated for a series of organic molecules at the HF/l-cc-pVDZ/$|\mathbf{B}| = 10^{-3}\,B_0$ level of theory. We divide the individual error by the magnetic field strength and the number of electrons to get an estimate for the screening error per electron. In (b) we use the opacity to differentiate between the maximum (transparent) and average (not transparent) error.
}
\label{fig2}
\end{figure*}

Let us now turn to the Mulliken approximations to the Berry curvature introduced in Section~\ref{Mulliken}. In Fig.~\ref{fig2}, we compare their errors with respect to the exact Berry curvature. The M2 approximation, based on Mulliken overlap populations, performs best in most of the investigated cases followed by the M3 approximation, which takes into account the changes of atomic dipole moments with respect to the nuclear coordinates. The simplest M1 approximation, including only the Mulliken atomic charges, performs worst. These trends are consistent across all three error measures ($\epsilon_\mathrm{tot}^\mathrm{MX}$, $\epsilon_\mathrm{avg}^\mathrm{MX}$, and $\epsilon_\mathrm{max}^\mathrm{MX}$).

The overall error [$\epsilon_\mathrm{tot}^\mathrm{MX}$, see Fig.~\ref{fig2}(a)] suggests that the M2 Berry curvature captures more than 90\% of the screening of H$_2$, CH$_4$, NH$_3$, H$_2$O, and FH correctly. This is remarkable considering the simplicity of the approach. With an error of about 70\%, the description of LiH and BH$_3$ is worse. This failure may be due to the challenging electronic structure of these low-valent molecules. The M3 and M1 models capture, on average, about 80\% and 70\% of the screening, respectively. While this was expected for the simple M1 approach, it is disappointing for the M3 approach, which requires the same computational effort as the exact Berry curvature. In general, the M1 model seems to perform best when the off-diagonal matrices of the Berry curvature are close to zero -- for example, in FH and H$_2$O.

\begin{figure*}
\centering
\begin{tabular}{ll}
(a) & (b) \\
\includegraphics[width=0.48\textwidth]{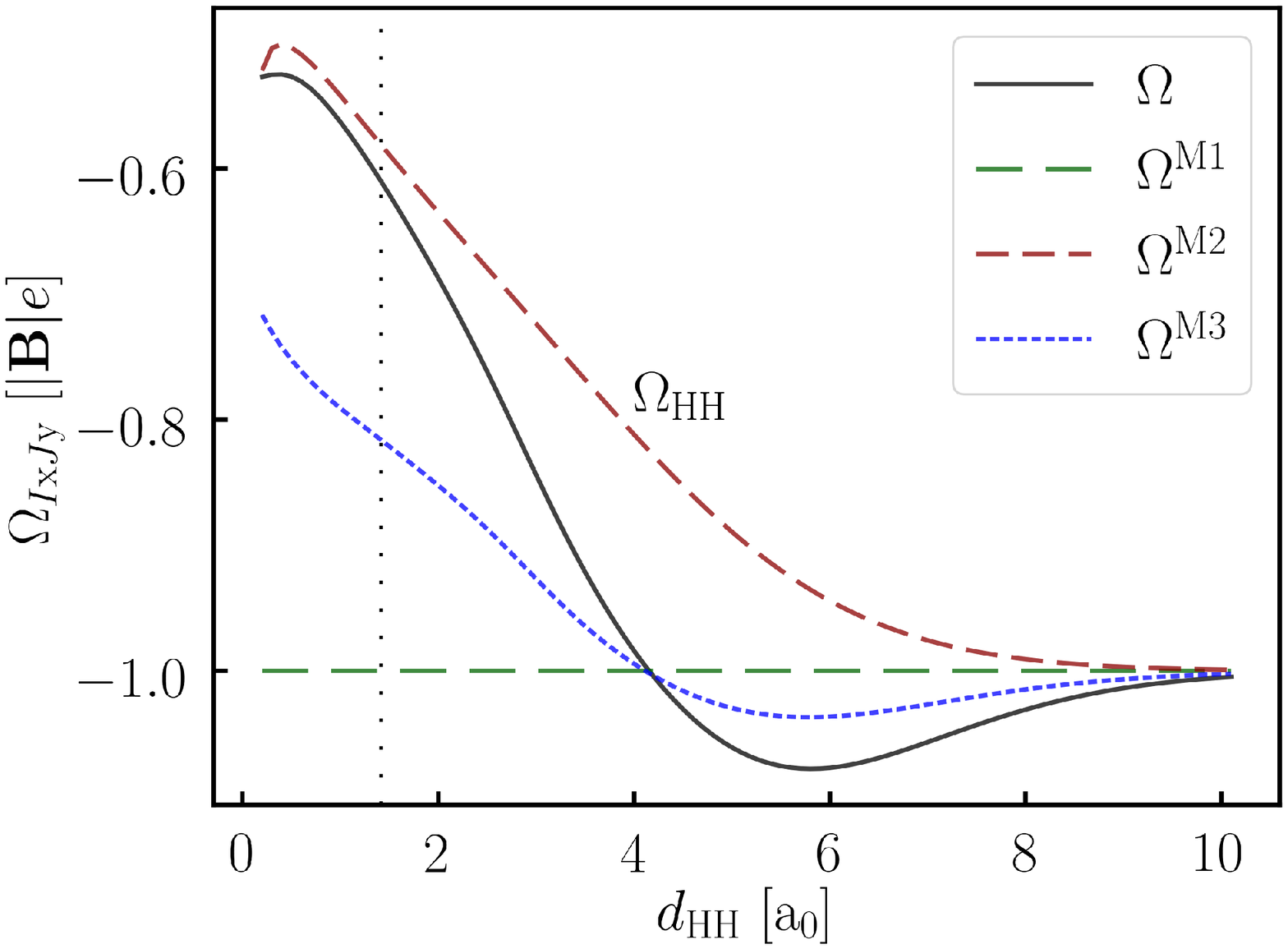} &
\includegraphics[width=0.48\textwidth]{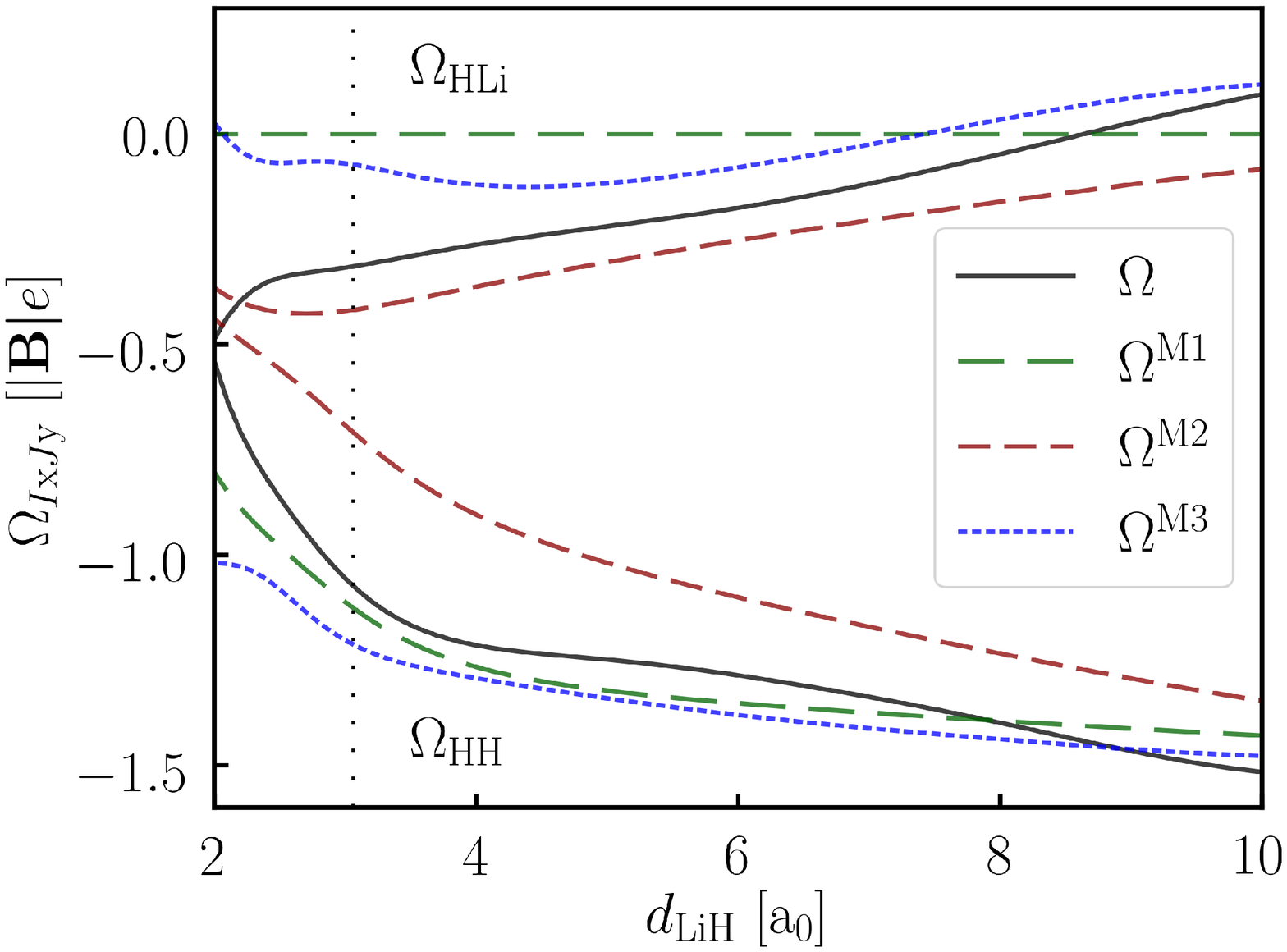} 
\end{tabular}
\caption{Components of the Berry curvatures [$\theom{}{}{d_\mathrm{XH}}$]  and their approximations [$\theoms{}{}{d_\mathrm{XH}}{MX}$, see eqs.~\eqref{mul_004}, \eqref{mul_008}, and \eqref{mul_013}] of H$_2$ (a) and LiH (b) at different bond lengths ($d_\mathrm{XH}$). All calculations were performed at the HF/l-cc-pVDZ/$|\mathbf{B}| = 10^{-3}\,B_0$ level of theory with $\mathbf{B}$ being perpendicular to the bonding axis. The equilibrium bond length is indicated by the vertical doted line.}
\label{fig3}
\end{figure*}

So far, we have only considered the Mulliken approximations to the Berry curvature at the equilibrium geometry. To investigate their behavior with  changes in the nuclear geometry, we have calculated the exact and approximate Berry curvatures of H$_2$ and LiH at different bond lengths ($d$); see Fig.~\ref{fig3}. As discussed in ref.~\onlinecite{Culpitt2021}, the limits $d \rightarrow 0/\infty$ correspond to the one atom/separate atoms limit. Whereas the M2 model reproduces both the exact united-atom and dissociation limits well, the M1 and M3 models are exact only in the dissociation limit. On the other hand, the M3 model is the only model that correctly predicts the superscreening [$\theomel{\mathrm{Hx}}{\mathrm{Hy}}{d_\mathrm{HH}} < -1$] of H$_2$ and the antiscreening [$\theomel{\mathrm{Hx}}{\mathrm{Liy}}{d_\mathrm{HLi}} > 0$] of LiH, suggesting that, even though the M3 model performs worse than the M2 model for the  $\epsilon^\mathrm{MX}$ metrics, its structure [see eq.~\eqref{mul_013}] is more flexible and able to recover more features of the exact Berry curvature.

As a final point, we stress that the approximations M1, M2, and M3 can also be determined from calculations without London orbitals. In our examples, with a relatively weak magnetic field of $10^{-3}B_0$ and the gauge origin and center of mass at the origin $(0,0,0)$, the resulting approximate Berry curvatures are very close to those calculated with London orbitals. Although they are not translationally invariant and dependent on the gauge origin, they may be a useful alternative when calculations with London orbitals are not possible -- in such cases, however, the gauge origin and the basis set must be chosen with great care.

\section{Conclusion and Outlook}

In this work, we have (re)investigated properties of the Berry connection and curvature that are crucial for the right physical behavior of the system in magnetic fields while showing their link to the screening of the nuclei by the electrons. It was demonstrated theoretically and via example calculations that these features are related to the magnetic-translational sum rule and a direct result of the translational behavior of the exact electronic wave function which is fully captured by the gauge factor used in finite London orbital basis sets. Since this can only be reproduced by regular basis sets in the complete basis-set limit, the use of London orbitals is essential when calculating these properties. 

Based on our derivations, we were able to establish a series of Mulliken approximations to the Berry curvature (M1, M2, and M3), which recover the properties of the exact Berry curvature, in particular, the correct center-of-mass motion during dynamics. While the M2 approximation performed best in our test cases, recovering approximately 90\% of the exact Berry curvature, future studies will focus on improvements of the M3 approach which seems to capture more features of the electronic screening process. Additionally, we are planning to test these approximations in molecular dynamics simulations in magnetic fields.

\clearpage
\section*{Data Availability}

The data that support the findings of this study are available within the article.

\section*{Acknowledgments}

This work was supported by the Research Council of Norway through ‘‘Magnetic Chemistry’’ Grant No.\,287950 and CoE Hylleraas Centre for Quantum Molecular Sciences Grant No.\,262695. This work has also received support from the Norwegian Supercomputing Program (NOTUR) through a grant of computer time (Grant No.\ NN4654K).

\section*{References}


%

\end{document}